# Energy Transfer from Individual Semiconductor Nanocrystals to Graphene

Zheyuan Chen[1][†], Stéphane Berciaud[1,2][†], Colin Nuckolls[1], Tony F. Heinz[2] and Louis E. Brus[1]*

[1]Department of Chemistry, Columbia University, New York, NY 10027, USA.

[2]Departments of Physics and Electrical Engineering, Columbia University, New York, NY 10027, USA.

†: these authors contributed equally to this work

* e-mail: leb26@columbia.edu

## Abstract

Energy transfer from photoexcited zero-dimensional systems to metallic systems plays a prominent role in modern day materials science. A situation of particular interest concerns the interaction between a photoexcited dipole and an atomically thin metal. The recent discovery of graphene layers permits investigation of this phenomenon. Here we report a study of fluorescence from individual CdSe/ZnS nanocrystals in contact with single- and few-layer graphene sheets. The rate of energy transfer is determined from the strong quenching of the nanocrystal fluorescence. For single-layer graphene, we find a rate of ~ 4ns$^{-1}$, in agreement with a model based on the dipole approximation and a tight-binding description of graphene. This rate increases significantly with the number of graphene layers, before approaching the bulk limit. Our study quantifies energy transfer to and fluorescence quenching by graphene, critical properties for novel applications in photovoltaic devices and as a molecular ruler.

## Introduction

Metallic surfaces are known to quench the fluorescence from nearby photoexcited dipoles through resonant energy transfer [1,2]. On the other hand, no energy transfer is expected when a dipole is placed in the vicinity of a transparent insulating surface. Graphene [3-5], as an atomically thin and nearly transparent semimetal represents an intermediate case of both fundamental and practical interest. Indeed, single-layer graphene (SLG) possesses extremely high carrier mobility [6], while absorbing only ~ 2% of incoming light, independent of wavelength across the visible spectrum [7, 8]. These properties make graphene an excellent candidate for solar cell electrodes [9] and other applications in photonics. Here we examine the interaction of the 2-dimensional graphene system with another model nanoscale system, that of 0-dimensional semiconductor nanocrystals. Such nanocrystals have broad and size-tunable absorption [10], and high photostability [11], which make them promising systems for diverse optical applications, including the light-harvesting material in photovoltaic cells [12-14].

Resonant energy transfer from nanocrystals to single and few-layer graphene is expected to occur, since these systems exhibit broad absorption across the visible spectral range. SLG, for example, is characterized by a linear band dispersion around the corners of its Brillouin zone (K and K' points) [5] and a nearly constant optical absorption. Near graphene, electronically excited species, such as semiconductor nanocrystals, can thus be quenched by resonant energy transfer, exciting electron-hole pairs in the semimetal [1]. Whether this rate is significant compared with the natural radiative decay is, however, presently unknown. Photoexcited semiconductor nanocrystals can also decay by a competing process of charge transfer to the graphene substrate. Photoinduced electron transfer to graphene would produce charged nanocrystals, which are understood to be responsible for the "off" periods in fluorescence blinking [15, 16]. Our measurements of core/shell CdSe/ZnSe nanocrystals adsorbed on single and few-layer graphene (FLG) also explore this potential decay channel.

## Results and Discussion

Graphene layers were deposited on quartz substrates by mechanical exfoliation [3] of kish graphite. Isolated CdSe/ZnS nanocrystals were then spun cast onto the samples. (See Methods for details.) Fluorescence from individual nanocrystals could be observed for nanocrystals located both on the bare quartz substrate and on a graphene layer (Figure 1). Strong fluorescence quenching was observed for particles deposited on graphene sheets compared to the bare substrate. The integrated fluorescence intensities varied significantly from nanocrystal to nanocrystal, on both quartz and graphene. We first calculated the average quenching factor $\rho = \frac{I^Q}{I^G}$, where $I^Q$ and $I^G$ are the fluorescence intensities (expressed in emitted photons per unit time) on quartz and on graphene, respectively. Each isolated diffraction limited fluorescence spot was assigned to an individual nanocrystal and fit to 2D Gaussian profile. Statistical distributions of the integrated intensities were constructed separately for both populations of nanocrystals on quartz and on graphene (Figure 2). The widths of the distributions show a considerable inhomogeneity. The average intensities in Figures 2c and 2d give a quenching factor of ~ 25 for SLG.

Different "blinking" behavior is observed for nanocrystals on quartz and on graphene (Figure 3). On quartz, long "off" periods occur; these are not observed on graphene. Different blinking behavior leads to different integrated intensities from one nanocrystal to the next, which complicates our quantitative measurement of quenching. Blinking is known to depend upon both the laser intensity[18] and on the nature of the underlying substrate [17-21]. However, nanocrystal fluorescence during the "on" period is known to have a relatively constant radiative rate [22] and near unity quantum yield [23]. Thus, in order to remove the effect of blinking, we used the following procedure to calculate the comparative intensities during the "on" periods only. In order to remain in the linear regime, a low laser excitation intensity of ~50 W/cm$^2$ was used to probe nanocrystals on quartz. A much shorter nanocrystal excited-state lifetime exists on SLG (Figure 2). We therefore used higher excitation intensity (~1500 W/cm$^2$) for nanocrystals on graphene, but with the same binning time (10ms) for recording the fluorescence emission. The integrated fluorescence signals from nanocrystals on graphene still show a linear relationship with laser intensity at this high value, indicating that the dependence of blinking behavior on excitation intensity is negligible. On quartz the "on" and "off" periods lead to a familiar bimodal distribution of fluorescence intensities [16] (Fig 3b). From a collection of more than 160 time traces on quartz, we found an average ratio of the "on" period $T_{on}$ to the integration time $T$ of 0.34. Variations in $T_{on}$ for different nanocrystals are chiefly responsible for the broad distribution shown in Figure 2d.

In contrast, fluctuations in the fluorescence intensity from nanocrystals on graphene are dramatically reduced. The fluorescence time traces yield a single-modal distribution of intensities (Figure 3b). This suppression of blinking suggests that the fluorescence quenching rate is significantly faster than the photoexcited electron trapping rate responsible for the "off" state. Most of the integration time is "on" for nanocrystals on graphene, and thus on graphene $T_{on}$ is approximated as $T$ in Figure 3a. The measured quenching factors were therefore corrected to account for the different "on" fractions, yielding $\rho \approx 80$.

We believe the quenching process, decreasing the nanocrystal quantum yield during the "on" periods, is resonant energy transfer and not electron transfer to graphene. Photoinduced electron transfer from core/shell nanocrystals to doped silicon substrates with a thin surface oxide, and to HOPG, has been studied by Electron Force Microscopy [24]. The rates were quite slow; such charge transfer would be negligible under our conditions of excitation intensity and integration time. In contrast, excited-state resonant energy transfer to graphene is predicted to be efficient. We express the corrected steady-state quenching factor $\rho$ (the inverse of the fluorescence quantum yield) in terms of the dipole radiative decay rate $\gamma_{rad}$ and non-radiative energy transfer rate $\gamma_{ET}$ : $\rho = \frac{\gamma_{rad} + \gamma_{ET}}{\gamma_{rad}}$.

We neglect any effects of optical reflection from graphene and also assume the nanocrystal fluorescence quantum yield in the "on" state is unity in the absence of graphene [23]. Using standard theoretical expressions for $\gamma_{rad}$ and the theoretical resonant energy transfer rate from an emitting dipole to the π electron system of SLG (approximated within the tight-binding model [25, 26]), we obtain (see Supporting Information):

$$\rho(1L) = \frac{\pi}{16} \times \frac{\alpha}{\varepsilon^{5/2}} \times \left(\frac{c}{v_F}\right)^4 \times I(z) + 1, \quad I(z) = \int_0^1 dt e^{-\frac{2\Delta Ezt}{\hbar v_F}} \frac{t^3}{\sqrt{1-t^2}} \quad \text{(equation 1)}$$

Here α is the fine structure constant, $\varepsilon$ is the dielectric constant of the surrounding medium, $c$ is the speed of light in vacuum, z is the distance from the nanocrystal center to the graphene plane, $\Delta E = 1.9\,\text{eV}$ is the energy of the emitted photons, and

$v_F = 1 \times 10^6 \, \text{ms}^{-1}$ is the Fermi velocity in SLG [5]. We take $\varepsilon$ to be that of the usual coating ligand trioctylphosphine oxide ($\varepsilon = 2.6$).

To our knowledge there is no theoretical expression for the corresponding energy transfer rate in few-layer graphene. Since the interactions between the layers of graphene are relatively weak [5] and we are concerned with excitations in the visible spectral range, we approximate the FLG system simply as a stack of decoupled single-layer graphene sheets. Each layer is treated as an independent energy transfer channel, separated from other layers by the graphite spacing of $\delta = 0.34$ nm. The dielectric screening from upper-layers of a FLG sample is assumed to be unchanged from that of the nanocrystal ligands. The quenching factor for FLG of n-layer thickness is then given by

$$\rho(nL) = \frac{\pi}{16} \times \frac{\alpha}{\varepsilon^{5/2}} \times \left(\frac{c}{v_F}\right)^4 \times \sum_{i=1}^{n} I(z_i) + 1, \quad \text{(equation 2)}$$

where $z_i = z_1 + (i-1)\delta$ is the distance from the nanocrystal center to the $i^{\text{th}}$ graphene layer.

A critical parameter in the model is the position of the nanocrystals with respect to the underlying graphene sheets. We measured this height distribution using nanocrystals dispersed on highly-oriented pyrolitic graphite (HOPG) by tapping-mode atomic-force microscopy (see Supporting Information, Figures S3 and S4). The average height of the top of the nanocrystals was found to be 6.1 nm; thus the average distance from the nanocrystal center to graphene is taken to be $z_1 \approx 3.05$ nm. From ref. [26], the theoretical distance ($z$) dependence of the dipole energy transfer rate to graphene is $z^{-4}$. As a result, smaller nanocrystals with lower $z_1$ should show larger $\rho$, and larger nanocrystals with greater $z_1$ should show smaller $\rho$. We do in fact observe a distribution of integrated fluorescence intensities for nanocrystals on graphene (Fig. 2b). For SLG we calculated the relative number of emitted photons from each part of the height distribution using equation 1. We found that the total number of emitted photons over the distribution was essentially the same as calculated using the average distance. Thus, in Figure 4 we compare data with theory using the average distance of $z_1 \approx 3.05$ nm.

The experimental and theoretical quenching factors $\rho$ are shown in Figure 4. The factors of 70 for SLG and ~115 for bilayer graphene are in good agreement with the dipole energy transfer theory in equation 2. Considering a typical radiative rate [19, 22] $\gamma_{rad} \sim 5\times10^7 s^{-1}$, we estimate $\gamma_{ET} \sim 4\times10^9 s^{-1}$ for SLG. The nanocrystal lifetime on graphene is about 250 ps. Interestingly, this value is similar to the reported near 200 ps$^{-1}$ lifetime of slightly smaller nanocrystals emitting at 620 nm on Au surfaces [18]. We note that in the case of bulk metals, surface roughness is known to cause dramatic modifications in the absorption and radiative decay rates, yielding either fluorescence enhancement or quenching [18]. In the case of atomically thin surfaces like graphene, such effects can be neglected so that a comparison of the fluorescence intensities is equivalent to a comparison of the excited-state lifetime. It is remarkable that nanocrystals on SLG, which only absorbs about 2% of incident light, have roughly the same lifetime as on flat Au metal.

The experimental fluorescence quenching factor $\rho$ increases with number of layers of the graphene sample, but is not in quantitative agreement with the model. This simple model should increasingly fail as the thickness increases, since it neglects attenuation and reflection of the emitting dipole near field in the top several layers for thick graphene samples [2]. For bulk graphite the measured $\rho$ is about 600, while the model calculated $\rho$ is only about 250. In the bulk limit, we can alternatively calculate the expected quenching $\rho$ using the well-known energy transfer theory for flat bulk materials based on the dielectric response of the medium [1] (see Supporting Information). This theory gives a quenching $\rho$ of 607, close to our measured value.

## Conclusions

We have demonstrated efficient energy transfer from individual CdSe/ZnS nanocrystals to single- and few-layer graphene. Our analysis corrects for the differing blinking kinetics observed on quartz and on graphene substrates. The fluorescence intensity of single nanocrystals is quenched by a factor of ~ 70 on single-layer graphene, in agreement with resonant energy transfer theory. The quenching efficiency increases with layer number. Resonant energy transfer is much faster than photoexcited electron transfer for hydrocarbon ligand coated, CdSe/ZnS core/shell nanocrystals adsorbed on graphene. How might one change the relative rates of electron transfer and energy transfer for solar energy applications? The rate of electron transfer could be increased by strengthening the electronic coupling between nanocrystal and graphene through covalent bonding and by removal of the strongly insulating ZnS outer shell. The photochemical covalent functionalization of graphene has been recently demonstrated [27], making possible strong electronic coupling between nanocrystals and graphene. The Fermi energy of graphene can also be tuned by electrostatic [28, 29] or chemical doping [30] in order to increase the rate of electron transfer and/or decrease the rate of resonant energy transfer. In addition to possibilities for photovoltaic devices mentioned above, the relatively strong fluorescence quenching that we have observed for graphene sheets suggests another promising possibility: The use of graphene and semiconductor nanocrystals (or other fluorophores) as a molecular ruler in which nanometer-scale distances are determined by analysis of fluorescence quenching. In particular, owing to the predicted $d^{-4}$ scaling of the rate of energy transfer [25, 26], fluorescence quenching by graphene could be used to measure distances that cannot be reached using standard donor-acceptor pairs [31], for which energy transfer decreases sharply as $d^{-6}$. These are subjects for future research.

**Methods**

Graphene layers were deposited onto clean quartz substrates by mechanical exfoliation [3] of kish graphite (Covalent Materials Corp). The number of graphene layers was determined by both Raman spectroscopy [32] and optical reflection contrast measurements [33] (see Supporting Information). CdSe/ZnS core/shell nanocrystals (Qdot 655, Invitrogen Corp., Cat. No. Q21721MP) were spuncoat onto the substrate at low density (< 0.4 $\mu m^{-2}$). Nanocrystals were illuminated under ambient conditions by a 532-nm continuous-wave diode laser for 30s at low laser intensity (~50 W/cm$^2$). The fluorescence from individual nanocrystals was collected by an air objective (100X, NA=0.9), sent through an emission filter (655 ± 20 nm), and imaged onto a CCD array (Figure 1a). Graphene pieces were located under white light illumination. The average fluorescence intensities were corrected for the slight inhomogeinity of the laser beam profile.

## Acknowledgments

We would like to thank H. Liu, E. Rabani, K. F. Mak, and L. Malard for fruitful discussions. This work was supported the Department of Energy through the EFRC program (grant DE-SC00001085) and the Office of Basic Energy Sciences (grant DE FG02-98ER14861) and by the New York State NYSTAR program.

**Figures:**

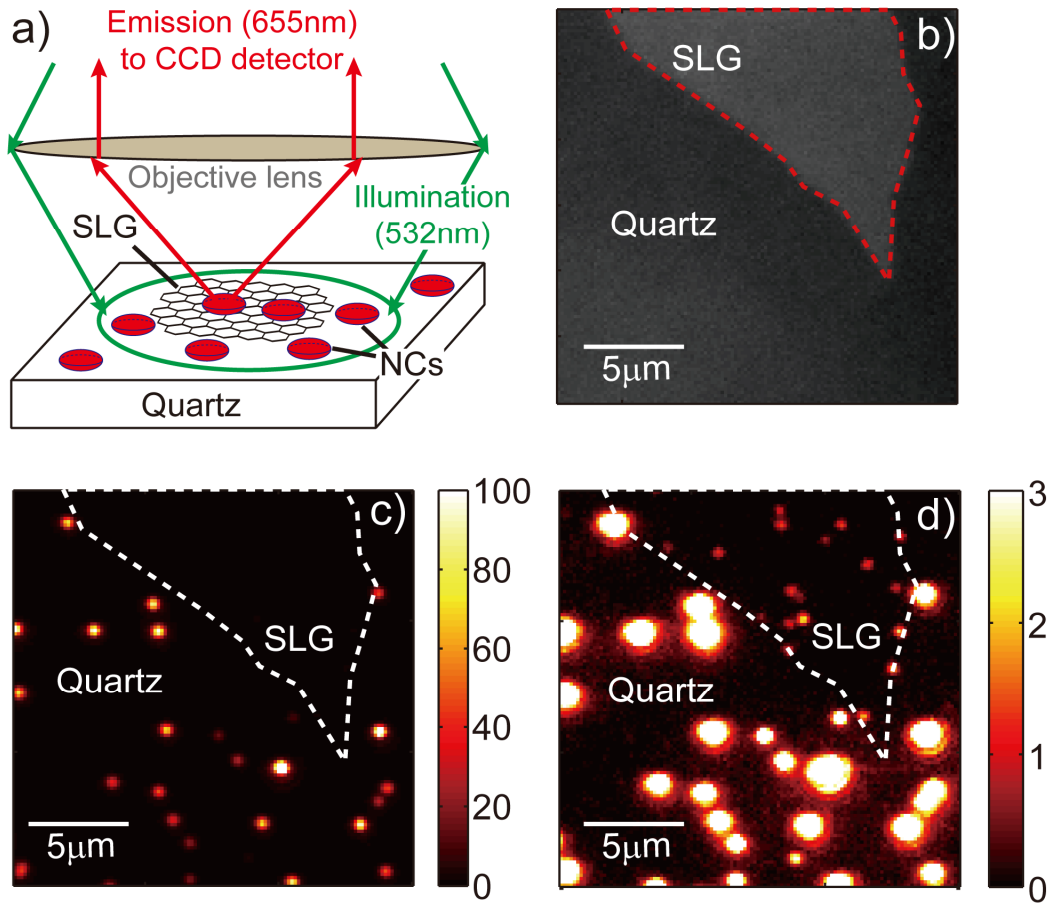

**Figure 1. Optical and fluorescence images of individual nanocrystals on single-layer graphene and on the quartz substrate.** a) Schematic diagram of our experimental setup. b) Optical reflectivity image in the emission range of our nanocrystals. c) Wide-field fluorescence image of individual CdSe/ZnS nanocrystals in the region shown in b). The color scale-bar indicates the number of emitted photons (in arbitrary units) integrated over 30s. d), Same as c), but in a color scale divided by a factor of 30 in order to show the emission from nanocrystals on a graphene monolayer.

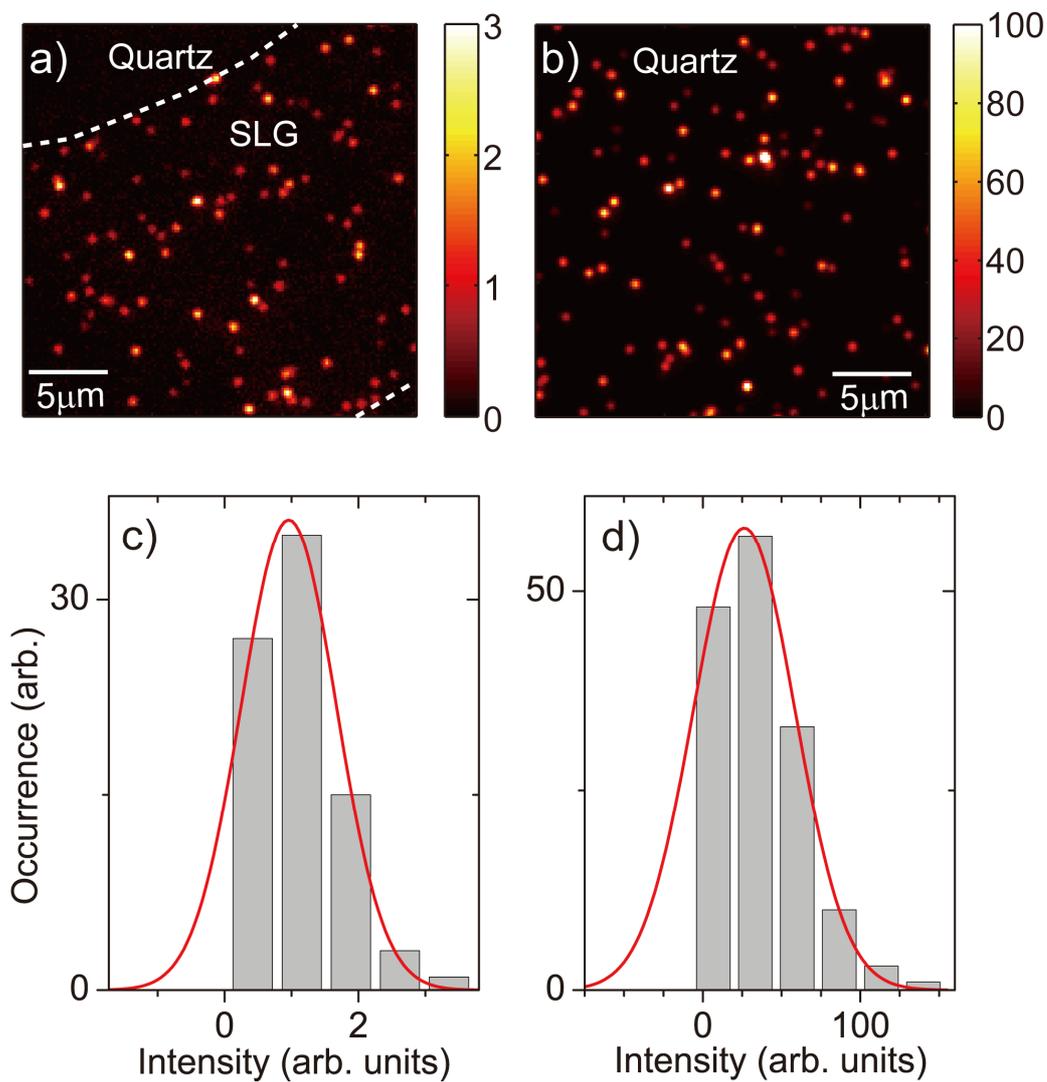

**Figure 2. Determination of the fluorescence quenching factor.** a) and c) Fluorescence images and corresponding histograms of the integrated fluorescence intensities for nanocrystals on a graphene monolayer, as compared to a reference taken on a quartz substrate (b) and d). The red curves in c) and d) show Gaussian fits to the histograms. The centers of the Gaussian profiles were used to calculate the average fluorescence quenching factors.

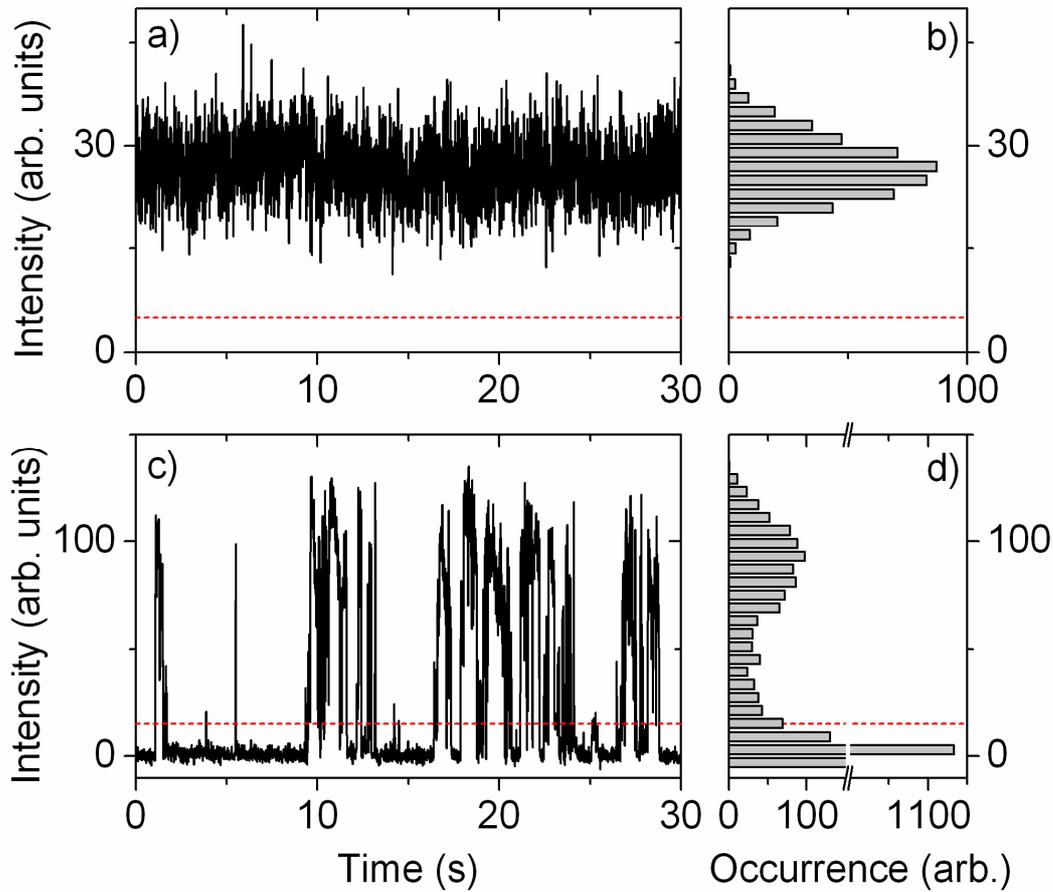

**Figure 3. Suppression of nanocrystal blinking on single-layer graphene.** a)**,** Fluorescence time traces from an individual nanocrystal lying on a graphene monolayer ($I_{Laser}$=1500W/cm$^2$) and b), on a quartz substrate ($I_{Laser}$=50W/cm$^2$). Both traces were acquired with a time bin of 10ms. The dashed horizontal lines indicate the intensity thresholds used to define the "on" and "off" states used in the text. c) and d), Histograms of the emission intensities corresponding to a) and b), respectively. After normalization for the laser excitation intensities, we deduce an average fluorescence quenching factor of ~75 between the "on" intensity measured on quartz and the intensity measured on graphene.

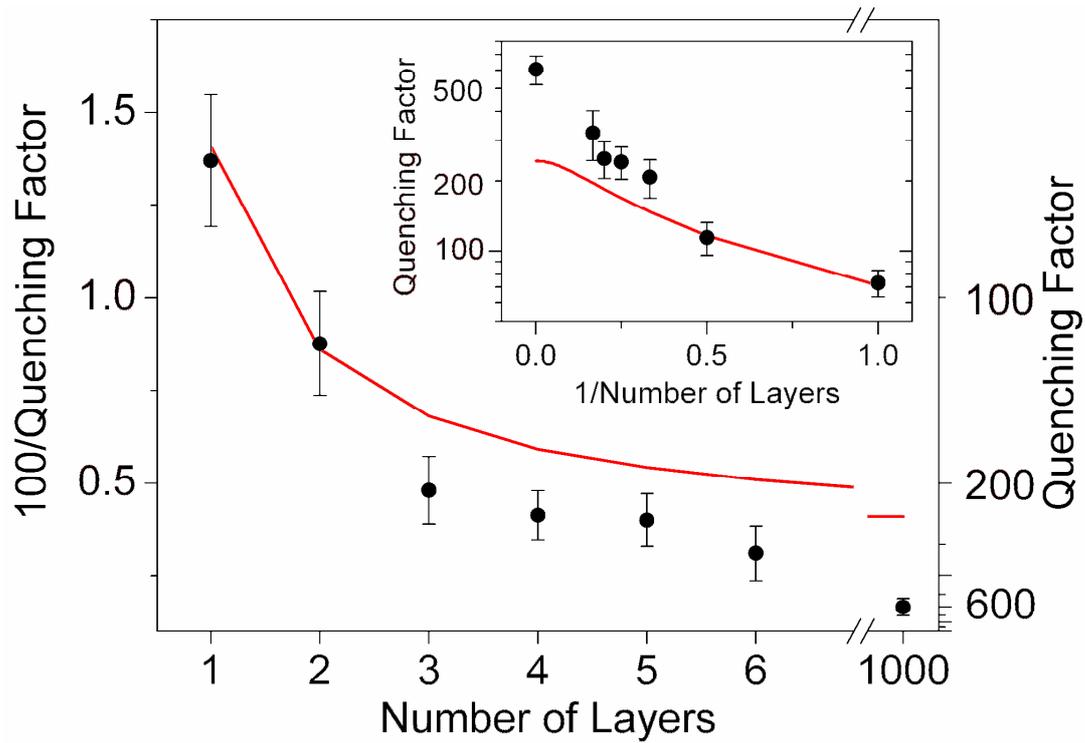

**Figure 4. Evolution of the fluorescence quenching factor with the number of graphene layers.** The black dots represent the quenching factors for single and few-layer graphene and for graphite determined from experiment, with the corresponding experimental uncertainties. The solid lines are the quenching factors calculated from the theory described in the text.